\documentstyle[nanosymp,epsf]{article}
\begin{document}
\title{Effects of accidental microconstriction on the quantized conductance in long wires}

\author{{\em A.A. Starikov}\dag \ddag, I.I.Yakimenko\dag,
K.-F. Berggren\dag, A.C. Graham\S, \\
K.J. Thomas\S, M. Pepper\S,  and M.Y. Simmons\S}

\affil{
\dag~ Department of Physics and Measurement Technology,\\
Link\"{o}ping University, S-581 83 Link\"{o}ping, Sweden\\
\ddag~ Kirensky Institute of Physics, 660036, Krasnoyarsk, Russia\\
\S Cavendish Laboratory, Madingley Road,\\
 Cambridge CB3 00HE, United Kingdom
}

\beginabstract
We have investigated the
conductance of long quantum wires formed in
$GaAs/Al_xGa_{1-x}As$ heterostructures.
Using realistic fluctuation potentials from donor layers
we have simulated numerically the conductance of four different
kinds of wires.
While ideal wires show perfect quantization,
potential fluctuations from random
donors may give rise to strong conductance oscillations and degradation
of the quantization plateaux.
Statistically there is always the possibility of having large fluctuations
in a sample that may effectively act as a microconstriction.
We therefore introduce microconstrictions
in the wires by occasional clustering of donors.
These microconstrictions are found to restore the quantized plateaux.
A similar effect is found for accidental lithographic inaccuracies.
\endabstract

\section*{Introduction}
Long quantum wires are  important objects from the view of
both technology and fundamental physics (see for example \cite{b1,b2,b3} and
references cited).
Theory predicts
that conductance through long clean wires is perfectly quantized with
much sharper  features than usual QPCs.
However, the practical manufacturing of perfect, high quality wires
is a delicate task.
Thus many factors may
influence the conductance  in real devices and may
cause deviations from theoretical predictions.
Potential fluctuations
from donors, small tolerance to possible inaccuracies in the lithography
of gates are some of these factors.
There could be two reasons for good quantization of the measured
conductance in a long
wire: either a superior quality of the wire or
effects not connected with
quantum properties of the entire wire, like local potential fluctuations
from donors or defects.
For example, in some samples we have observed quite well defined plateaux that
degrade on thermal cycling. Does it mean that the
potential landscape is changed
in this process in such way that an occasional microconstriction is removed?

In this work we discuss the conductance of long wires
in $GaAs/Al_xGa_{1-x}As$ heterostructures. We consider some accidental effects
that could produce well shaped quantized plateaux.
In particular, we study how donors
and lithographical deffects can produce QPC-like substructures and
change the shape of the conductance plateaux characteristics.

\section{Theoretical model and calculation}
Long quantum wires are often fabricated in modulation-doped
heterostructure with a patterned metallic top gate, see Fig. 1a.
\begin{figure}[h]
\centering
\begin{minipage}[t]{0.87\textwidth}

\begin{minipage}[t]{0.47\textwidth}
\leavevmode
%
\centering{\epsfxsize=\textwidth \epsfbox{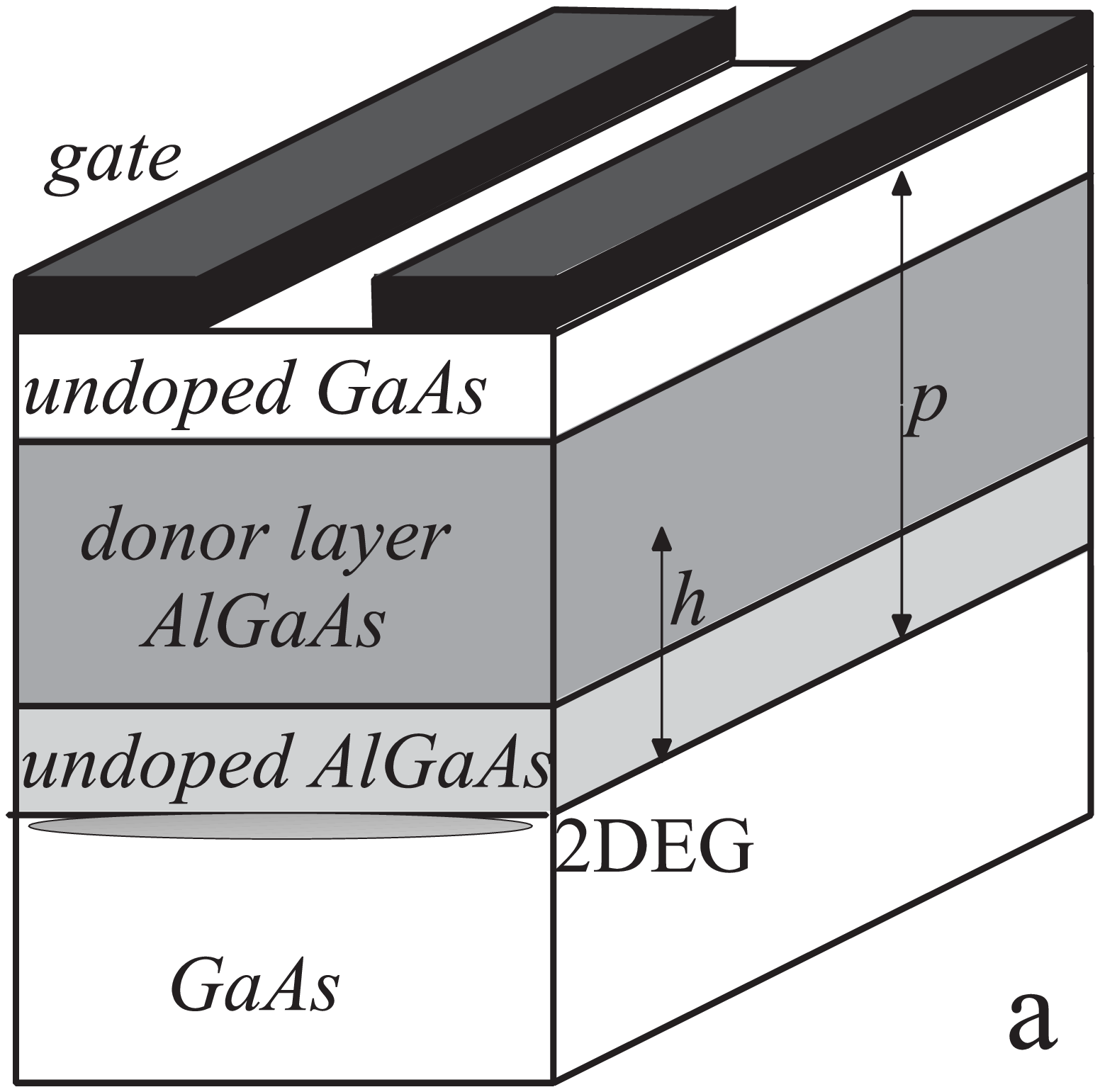}}
\end{minipage}
\hfill
\begin{minipage}[t]{0.47\textwidth}
\leavevmode
\centering{\epsfxsize=\textwidth \epsfbox{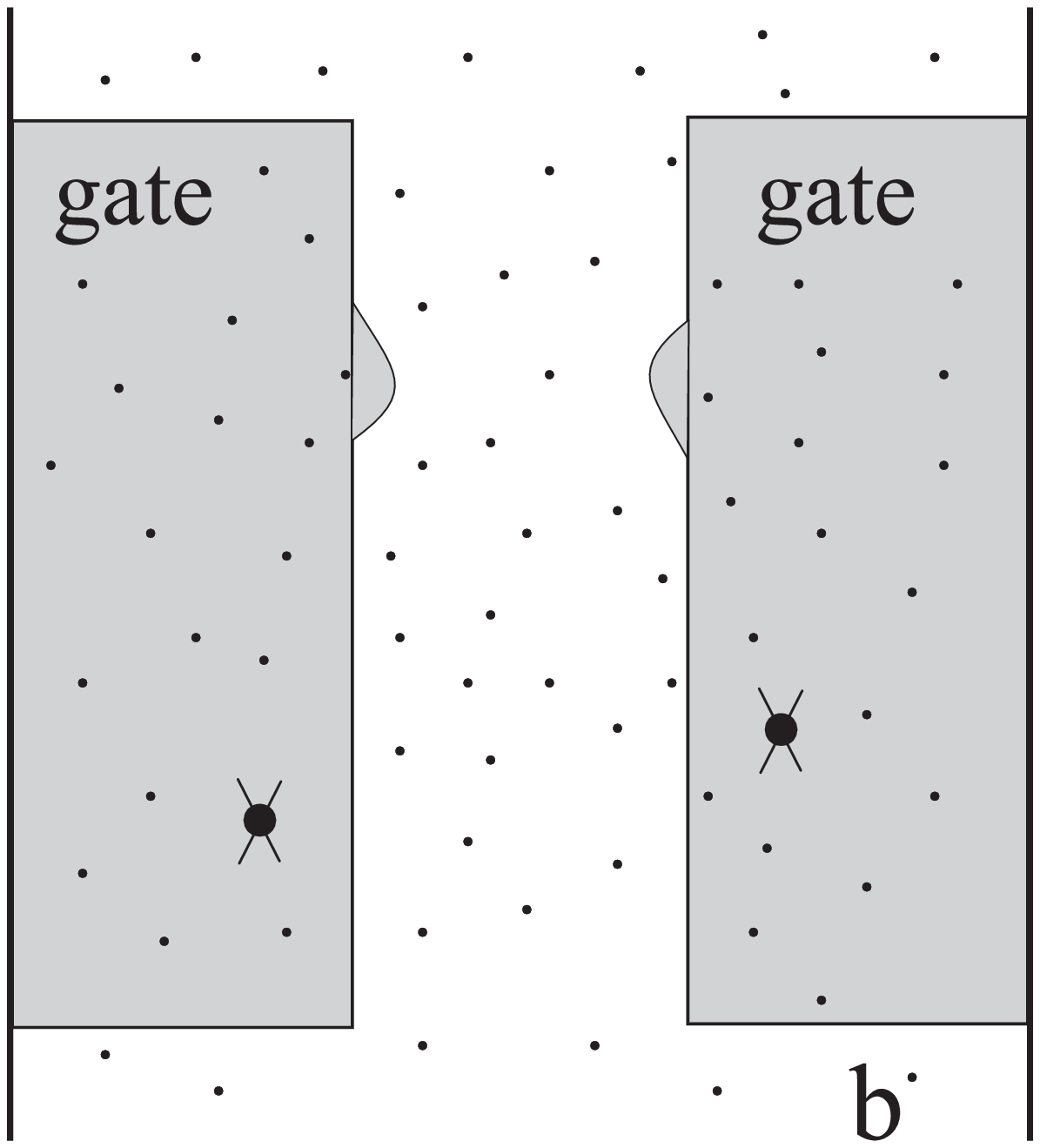}}
\end{minipage}
\caption{
a) Schematic representation of a device simulating a long quantum
wire;
b) Donors in the doped $AlGaAs$(dots), points of concentration
of donors (crosses), irregularities of the shape the gates. Corresponding
confinement potentials are smoothed.
}
\end{minipage}
\end{figure}
A negative voltage applied to the gate relative to the substrate
depletes the two-dimensional electron gas (2DEG) under the gated regions
and leaves a conducting wire in the split gate region.
The actual size of the wire  can be varied lithographically
by changing the geometry of the gate as well as by varying the
applied voltage. The confinement potential $U_g$
formed by the gate is determined by the well known
expression \cite{b4} :
\begin{equation}
U_g({\bf r},z)={{\it e} \over 2\pi}\int d{\bf r'} V_g({\bf r'},0)
{|z| \over (z^2+|{\bf r}-{\bf r'}|^2)^{3/2}}\:,
\end{equation}
where $V_g({\bf r'},0)$ is the potential on the gate and $z$ the distance
between  the gate and the 2DEG.
To investigate possible effects in real devices we introduce the potential
from random donors and mirror charges as described in \cite{b5} :
\begin{equation}
U_{rc}({\bf r},z)={{\it e}\over 4\pi\epsilon\epsilon_0}\sum_{i=1..N} \Big[{1 \over \sqrt{h^2+|{\bf r}-{\bf r'}|^2}}
-{1 \over \sqrt{(2p-h)^2+|{\bf r}-{\bf r'}|^2}}\Big]\: ,
\end{equation}
where the summation is over all donor positions in the
plane on distance $h$ from
the 2DEG. Mirror charges are generated by the surface at distance $p$ from
the 2DEG in Fig. 1a.

To find the electronic configuration and related properties of the
system we have solved the
Schr\"{o}dinger equation for electrons of a mass $m^*$ mapped on a square
lattice:
\begin{equation}
-{h^2\over2m^*} \nabla^2\Psi(x,y)+[U_g+U_{rc}]\Psi(x,y)=E_f
\Psi(x,y)\:.
\end{equation}
The zero-temperature conductance was calculated
as $G=(2{\it e}^2/h) Tr(TT^+)$
according to the B\"{u}ttiker formula; $T$ is the transmission matrix.
We use this one electron approach because we believe that brings out the
essential physics in the problem.

In the following we consider four variants of long wires:

\noindent a) a wire with uniform potential (ideal wire);

\noindent b) a wire with a fluctuating potential $U_{rc}$ to mimic a
real device.

\noindent We also consider two additional cases associated with possible
problems of manufacturing long wires. In both cases we assume
that there are background potential fluctuations
from random donors as in b). Thus we also have

\noindent c) a wire with clusters of donors. We simply let more than one donor
sit at the same point as shown in Fig. 1b.
The clusters represent statistical variations in donor concentration and/or
impurities or other imperfections from the growth of the heterostructure;

\noindent d) the top gate has deviations from the regular shape as shown
in Fig. 1b.
As a result an additional QPC microconstriction in the wire is formed.

We have studied wires formed by split gates as in Fig. 1.
The lithographic length is $4\:\mu m$ and the
gap between gates is $100\:nm$ wide.
Typical size of the numerical grid is 70x1000.

\section{Results and discussion}
The calculated conductance as a function of applied gate voltage is
show in Fig. 2. As indicated,
the conductance for an ideal wire (dotted line) shows well-shaped sharp stairs.
The dashed line on the same figure represents transport through the same
long wire with potential fluctuations  from random donors.
\begin{figure}[h]
\centering
\begin{minipage}[t]{0.8\textwidth}
\leavevmode
\centering{\epsfxsize=\textwidth \epsfbox{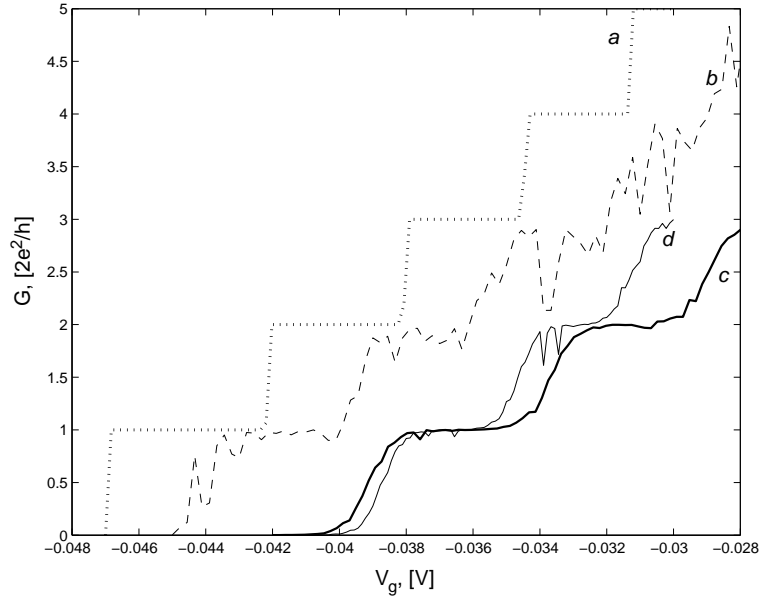}}
\caption{
Calculated zero-temperature conductance for long wires.
a) ideal wire; b) wire with fluctuated potential; c) microconstriction
substructure formed
by donor potential; d) microconstriction substructure from irregular gates.
}
\end{minipage}
\end{figure}
In this case the quantization is degraded  because of
 backscattering and interference effects.
Depending on the particular realization of random positions of donors
the  conductance oscillations differ from sample to sample.
The situation changes dramatically when we add two donors having larger charge,
that represent  local concentrations/statistical clusters of donors or other
imperfections.
This case is shown by the bold solid line. Donors situated on the different
sides of the channel may form a QPC-like structure which plays
a dominant  role in the transport process through the channel.
The case when a QPC-like structure is formed by
lithographic irregularities in the gates
looks similar.
However, the main difference between the two cases is the width of the plateaux
which is a result of the different mechanisms behind the formation
the QPC-like structures.
In the case of gate irregularities the potential is more responsive to the
applied voltage, it acts like a normal QPC (Fig. 2d.).
In contrast, when the narrowing is due to
donor concentrations/clusters {\it etc.}, variation of voltage does
not change the structure of local QPC-potential.
Instead the voltage only shifts the entire potential landscape.
%
In this case the plateaux are wider indicating less dependence on gate voltage
(Fig. 2c).
%

In conclusion, the microconstrictions in long quantum wires
may play a dominant role in transport processes.
Good quantization may be observed not
only in the case of well manufactured devices but also for imperfect
wires with microconstricitions.
Experimentally it may be a problem to recognize what kind of situation
we have at hand.
One possible way  is to compare wires with the same geometry.
Plateaux for QPC substructures are shifted with gate voltage and
have different shapes which may be analyzed by means of a local saddle-point
model.


\subsection*{Acknowledgments}
This work has been supported by the the Royal Swedish Academy of Sciences
(AAS) and the Engineering and Physical Sciences Research Council, UK (ACG).
KJT acknowledges support from Royal Society Research Fellowship, UK.

\end{document}